\documentclass[aps,prl,twocolumn,superscriptaddress,amsmath]{revtex4-1}

\usepackage{graphicx}

\begin{document}

%\preprint{}

\title{Long-range spin-polarized quasiparticle transport in mesoscopic Al superconductors with a Zeeman splitting}


\author{F. H\"ubler}
\affiliation{Institut f\"ur Nanotechnologie, Karlsruher Institut f\"ur Technologie, Karlsruhe, Germany}
\affiliation{Center for Functional Nanostructures, Karlsruher Institut f\"ur Technologie, Karlsruhe, Germany}
\affiliation{Institut f\"ur Festk\"orperphysik, Karlsruher Institut f\"ur Technologie, Karlsruhe, Germany}
\author{M. J. Wolf}
\affiliation{Institut f\"ur Nanotechnologie, Karlsruher Institut f\"ur Technologie, Karlsruhe, Germany}
\author{D. Beckmann}
\email[e-mail address: ]{detlef.beckmann@kit.edu}
\affiliation{Institut f\"ur Nanotechnologie, Karlsruher Institut f\"ur Technologie, Karlsruhe, Germany}
\affiliation{Center for Functional Nanostructures, Karlsruher Institut f\"ur Technologie, Karlsruhe, Germany}
\author{H. v. L\"ohneysen}
\affiliation{Center for Functional Nanostructures, Karlsruher Institut f\"ur Technologie, Karlsruhe, Germany}
\affiliation{Institut f\"ur Festk\"orperphysik, Karlsruher Institut f\"ur Technologie, Karlsruhe, Germany}
\affiliation{Physikalisches Institut, Karlsruher Institut f\"ur Technologie, Karlsruhe, Germany}

\date{\today}

\begin{abstract}
We report on nonlocal transport in multiterminal superconductor-ferromagnet structures, which were fabricated by means of e-beam lithography and shadow evaporation techniques. In the presence of a significant Zeeman splitting of the quasiparticle states, we find signatures of spin transport over distances of several $\mu$m, exceeding other length scales such as the coherence length, the normal-state spin-diffusion length, and the charge-imbalance length. The relaxation length of the spin signal shows a nearly linear increase with magnetic field, hinting at a freeze-out of relaxation by the Zeeman splitting. We propose that the relaxation length is given by the recombination length of the quasiparticles rather than a renormalized spin-diffusion length.
\end{abstract}

% 73.23.-b 	Electronic transport in mesoscopic systems
% 73.40.Gk = Tunneling
% 74.25.F- = Superconductors / Transport properties
% 74.40.Gh = Nonequilibrium superconductivity
% 74.55.+v 	Tunneling phenomena: single particle tunneling and STM
% 74.78.Na 	Mesoscopic and nanoscale systems 
%\keywords{}

\maketitle


The creation and and manipulation of spin-polarized currents forms the basis of spintronics applications \cite{zutic2004}. One key ingredient is the ability to transport spin currents over mesoscopic length scales, which are usually limited by spin-flip or spin-orbit scattering processes. Superconductors are particularly interesting for spin injection experiments due to the possibility to create almost 100~\% spin polarization \cite{giazotto2008}, enhanced spin relaxation times \cite{yafet1983,*yamashita2002,*morten2004,*morten2005}, and the separation of spin and charge degrees of freedom \cite{zhao1995}. Only few experiments on spin injection into superconductors have been reported so far (see \cite{johnson1994,poli2008,yang2010} and references therein). Both anomalously short \cite{yang2010} and anomalously long \cite{poli2008} relaxation times as compared to the normal state have been reported. Here, we report on investigations of spin transport in superconductors in the regime of large Zeeman splitting. In this 
regime, a current spin polarization of 100~\% can be achieved \cite{giazotto2008}, which is implicitly assumed in the classic experiments on spin polarized tunneling in high magnetic fields \cite{tedrow1971,*meservey1994}. 
We study in detail the diffusion of spin-polarized quasiparticles by using nonlocal detection with ferromagnetic electrodes as a function of contact distance, temperature and magnetic field, and present evidence for spin transport over surprisingly long distances.

\begin{figure}[bt]
\includegraphics[width=\columnwidth]{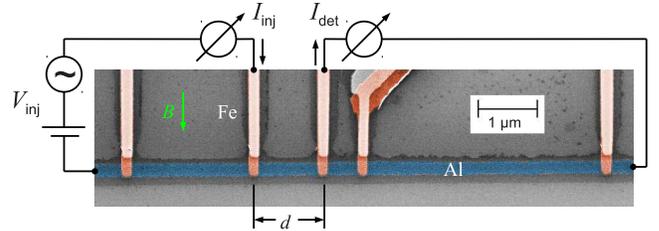}
\caption{\label{fig_sample}
(color online) Scanning electron microscopy image of a sample together with the measurement scheme with the injection (inj) and detection (det) circuits.}
\end{figure}

Our samples were fabricated by e-beam lithography and shadow evaporation techniques. They consist of a thin superconducting (S) aluminum strip of thickness $t_\mathrm{Al}\approx10-15$~nm, which was oxidized \textit{in situ} to form an insulating (I) tunnel barrier before being overlapped by several ferromagnetic (F) iron contacts ($t_\mathrm{Fe}\approx15-25$~nm). In addition a copper layer ($t_\mathrm{Cu}\approx30$~nm) was evaporated under a third angle to reduce the resistance of the iron leads. Consistent results were obtained from nine samples of slightly different designs. We focus here on one sample (labeled FISIF) for which the most complete data set was recorded, and data from a reference sample with normal-metal (N) copper contacts, but otherwise similar parameters (labeled  NISIN). Figure~\ref{fig_sample} shows a scanning electron microscopy image of the FISIF sample, together with the experimental scheme. The sample has five contacts, spanning contact distances $d$ from $0.5$ to $8~\mathrm{\mu m}$.

The local ($g_\mathrm{loc}=dI_\mathrm{inj}/dV_\mathrm{inj}$) and nonlocal ($g_\mathrm{nl}=dI_\mathrm{det}/dV_\mathrm{inj}$) differential conductance for different contact pairs was measured by standard lock-in techniques in a setup described elsewhere \cite{huebler2010,huebler2012}. Measurements were performed in the superconducting state of aluminum at temperatures down to $T=50~\mathrm{mK}$, and with an in-plane magnetic field $B$ applied parallel to the ferromagnetic wires. For all data shown here the magnetization of the iron wires is aligned parallel to the magnetic field. We also performed nonlocal spin-valve experiments in the normal state at $T=4.2~\mathrm{K}$ (not shown), from which the normal-state spin diffusion length $\lambda_\mathrm{N}=370\pm10~\mathrm{nm}$ and the spin polarization of the tunnel conductance  $P=(G_\downarrow-G_\uparrow)/(G_\downarrow+G_\uparrow)=0.19\pm0.05$  were obtained. Here, $G_{\uparrow,\downarrow}$ are the junction conductances for each spin.

\begin{figure}[bt]
\includegraphics[width=\columnwidth]{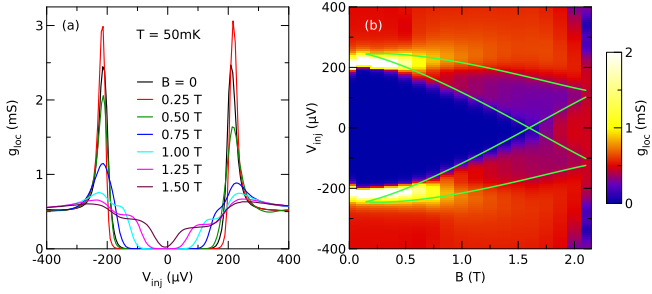}
\caption{\label{fig_local} 
(color online) (a) Local differential conductance $g_\mathrm{loc}=dI_\mathrm{inj}/dV_\mathrm{inj}$ of one junction as a function of injector bias $V_\mathrm{inj}$ for different applied magnetic fields $B$. (b) The same data plotted on a color scale. The lines indicate the regions where a single spin band dominates conductance.}
\end{figure}

Figure~\ref{fig_local}(a) shows the local differential conductance of one contact as a function of the injection bias voltage $V_\mathrm{inj}$ for different applied magnetic fields $B$ at $T=50~\mathrm{mK}$. For small $B$ pronounced gap features at $V\approx\pm205~\mu$V are observed as well as a negligible subgap conductance. Upon increasing the magnetic field, the gap features broaden due to orbital pair breaking, and for $B>0.5~\mathrm{T}$ the Zeeman splitting is seen. We describe our data with the standard model of high-field tunneling \cite{tedrow1971,*meservey1994} to obtain the spin-dependent density of states $n_\sigma(E)$, where $\sigma=\pm 1$ stands for spin up and down, respectively.
From  $n_\sigma(E)$, we calculate the current for each spin
\begin{equation}
I_\sigma=\frac{G_\mathrm{N}}{2e}\int (1-\sigma P)n_\sigma(E)\left[f_0(E)-f_0(E+eV)\right]dE \label{equ_gperspin}
\end{equation}
where $G_\mathrm{N}=G_\downarrow+G_\uparrow$ is the normal-state junction conductance, and $f_0$ is the Fermi function. The total charge current is $I=I_\uparrow+I_\downarrow$, and the spin current is proportional to $I_s=I_\uparrow-I_\downarrow$. Fits of this model to the measured conductance spectra yield $G_\mathrm{N}$, the pair-breaking parameter $\Gamma$, and the spin-orbit scattering strength $b_\mathrm{so}$. Details of the fit procedure have been given previously \cite{huebler2010,huebler2012}. The spin polarization $P=0.19\pm0.01$ obtained from these fits is the same as obtained from the spin-valve experiments. The relatively small $P$ is typical of ultra-thin alumina tunnel barriers \cite{muenzenberg2004}. Figure~\ref{fig_local}(b) shows a contour plot of the complete dataset of the local conductance as a function of bias and magnetic field. The gap observed at $B=100~\mathrm{mT}$ is slightly larger than at zero applied field. We attribute this to the presence of stray fields of the ferromagnetic contacts. At higher fields, in the wedge-shaped regions indicated by the lines, a single spin band dominates conductance. 

\begin{figure}[tb]
\includegraphics[width=\columnwidth]{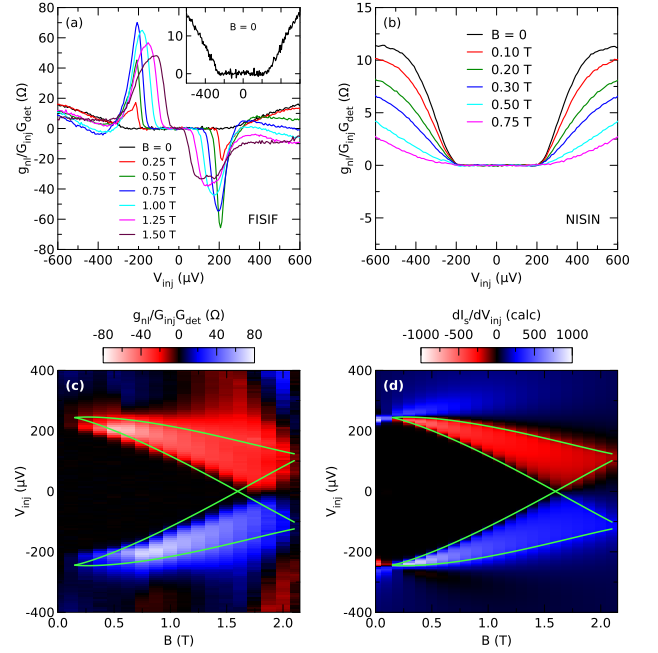}
\caption{\label{fig_nonlocal} 
(color online) Normalized nonlocal differential conductance $g_\mathrm{nl}/G_\mathrm{inj}G_\mathrm{det}$ as a function of injector bias $V_\mathrm{inj}$ for different applied magnetic fields $B$ for one pair of contacts (a), nonlocal conductance of a pair of contacts of the NISIN reference sample (b), the data from panel (a) plotted on a color scale (c), and calculated differential spin current (d).}
\end{figure}

Next, we focus on the nonlocal differential conductance. To eliminate the effect of small variations of the junction conductances, we plot the normalized nonlocal conductance $g_\mathrm{nl}/G_\mathrm{inj}G_\mathrm{det}$ throughout the paper. In Fig.~\ref{fig_nonlocal}(a) $g_\mathrm{nl}/G_\mathrm{inj}G_\mathrm{det}$ is displayed as a function of the applied bias voltage $V_\mathrm{inj}$ for different magnetic fields $B$ and a contact distance $d\approx1~\mathrm{\mu m}$. The data were measured simultaneously with the local conductance of Fig.~\ref{fig_local}(a), in the configuration shown in Fig.~\ref{fig_sample}. For comparison, we show data obtained from the NISIN reference sample in Fig.~\ref{fig_nonlocal}(b). At $B=0$, there is no conductance below the gap, and above the gap, both the FISIF and NISIN samples show a nearly linear increase due to charge imbalance \cite{huebler2010}. With increasing magnetic field, the charge imbalance signal decreases, as clearly seen for the NISIN sample. The FISIF sample shows a qualitatively different behavior: ($i$) in the bias range corresponding to the Zeeman splitting, a positive peak arises for $V_\mathrm{inj}<0$, and a negative peak for $V_\mathrm{inj}>0$; ($ii$) for higher bias $|V_\mathrm{inj}|\gtrsim 300~\mathrm{\mu V}$, an additional asymmetry evolves on top of the charge imbalance signal. While the observation ($i$) is systematic for all nine samples, ($ii$) was observed only in a few samples, whereas other samples showed the symmetric charge imbalance signal also seen in the NISIN sample at high bias. In the following we therefore only concentrate on the asymmetric peak features. Upon increasing the field, the peak heights increase gradually to their extremal values at $B\sim0.5-0.75$~T, before the peaks start to decline, broaden and move inwards, simultaneously. The positive peak (at negative bias) is slightly larger than the negative peak (at positive bias). Above the critical field $B_{c}\approx 2.15~\mathrm{T}$ the asymmetric features disappear and one finds a small bias-independent signal (not shown). 

Figure~\ref{fig_nonlocal}(c) represents the same data as in Fig.~\ref{fig_nonlocal}(a) as a contour plot with additional magnetic field steps. The lines are duplicated from  Fig.~\ref{fig_local}(b). As can be seen, the asymmetric conductance features are limited almost entirely to the magnetic-field and bias region of the Zeeman splitting. For comparison,  in Fig.~\ref{fig_nonlocal}(d) we show the differential spin current $dI_s/dV_\mathrm{inj}$ calculated using (\ref{equ_gperspin}) and parameters obtained from the fits to the local conductance of the injector.

\begin{figure}[bt]
\includegraphics[width=\columnwidth]{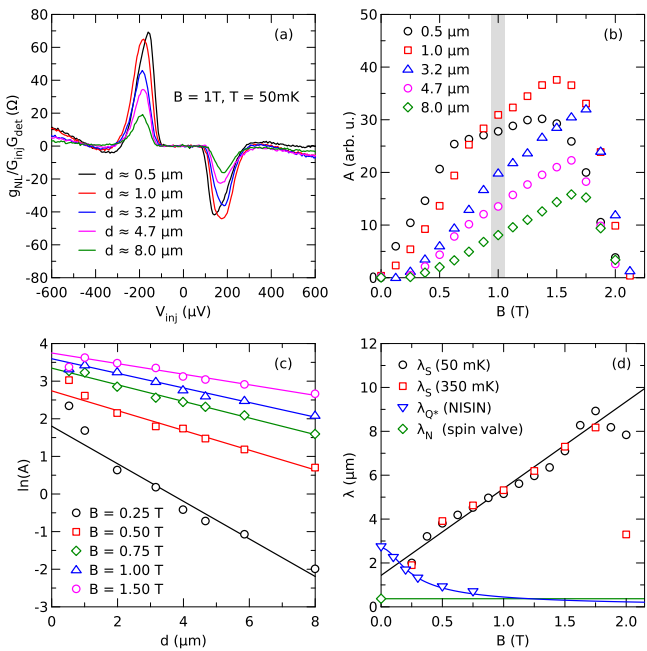}
\caption{\label{fig_dist}
(color online) (a) Normalized nonlocal differential conductance $g_\mathrm{nl}/G_\mathrm{inj}G_\mathrm{det}$ as a function of the bias voltage $V_\mathrm{inj}$ for different contact distances $d$ at fixed magnetic field $B$ (b) peak area $A$ (for the peak at $V_\mathrm{inj}<0$) as a function of $B$ for different $d$ (c) semi-logarithmic plot of $A$ as a function of $d$ for different $B$, together with fits to an exponential decay (d) relaxation length $\lambda_\mathrm{S}$ derived from these fits as a function of $B$, together with the charge imbalance length $\lambda_{Q^*}$ of the reference NISIN structure, and the normal-state spin diffusion length $\lambda_\mathrm{N}$ derived from nonlocal spin-valve experiments.}
\end{figure}

Fig.~\ref{fig_dist}(a) shows the normalized nonlocal conductance as a function of the bias voltage for different contact distances $d$ and one fixed magnetic field and temperature ($B=1$~T and $T=50$~mK). In general, increasing the distance between injector and detector decreases the amplitude of the signal, but does not affect its overall shape. Only the signal for the shortest distance $d=0.5~\mathrm{\mu m}$ slightly deviates in shape. 

We have also analyzed the heights and areas of both peaks as a function of magnetic field, temperature and contact distance. Height and area for both peaks yielded similar results, and we only show the peak area $A$ of the positive peak in the following. Fig.~\ref{fig_dist}(b) shows $A$ as a function of the applied field $B$ for different distances $d$ at $T=50~\mathrm{mK}$. $A$ rises monotonically with increasing magnetic field, until at $B\approx 1.5-1.75~\mathrm{T}$ it reaches a maximum and then rapidly goes back to zero. Again, this behavior is similar for all distances, with slight deviations at $d=0.5~\mathrm{\mu m}$.
 
Fig.~\ref{fig_dist}(c) shows the peak area as a function of contact distance on a semi-logarithmic scale for different magnetic fields. The solid lines are fits to an exponential decay, from which we obtain the spin relaxation length $\lambda_\mathrm{S}$ (see discussion). As can be seen, the quality of the fits is very good up to $B\approx 1.5~\mathrm{T}$, except for the shortest distance $d=0.5~\mathrm{\mu m}$, which was therefore excluded from the fits. At fields above $1.5~\mathrm{T}$, the quality of the fits declined (not shown).

Fig.~\ref{fig_dist}(d) shows $\lambda_\mathrm{S}$ as a function of the applied magnetic field $B$ for two different temperatures. It is about $2~\mathrm{\mu m}$ for small magnetic fields and then increases nearly linearly with $B$ up to around $7~\mathrm{\mu m}$ for $B\approx1.5~\mathrm{T}$.  At higher fields, $\lambda_\mathrm{S}$ appears to decline, but as mentioned above the fits were not very good in this region. $\lambda_\mathrm{S}$ is almost independent of temperature up to $350~\mathrm{mK}$. For comparison, we also show the charge-imbalance relaxation length $\lambda_{Q^*}$ obtained from the NISIN reference sample, and the spin-diffusion length $\lambda_\mathrm{N}$ from the normal-state spin-valve experiments.

\begin{table}[bt]
\begin{tabular}{lccc}
 & Operators & Probability & Electrons \\
 &           &             & added to S \\\hline
1 & $\gamma^\dagger_{\mathbf{k}\sigma}c_{\mathbf{q}\sigma}$ & 
$u^2_\mathbf{k}(1-f_{\mathbf{k}\sigma})
f_0(E_{\mathbf{k}\sigma}-eV)$ & +1 \\
2 & $\gamma_{\mathbf{k}\sigma}c_{\overline{\mathbf{q}}\overline{\sigma}}$ &
$v^2_\mathbf{k}f_{\mathbf{k}\sigma}
(1-f_0(E_{\mathbf{k}\sigma}+eV))$ & +1 \\
3 & $\gamma_{\mathbf{k}\sigma}c^\dagger_{\mathbf{q}\sigma}$ & 
$u^2_\mathbf{k}f_{\mathbf{k}\sigma}
(1-f_0(E_{\mathbf{k}\sigma}-eV))$ & -1 \\
4 & $\gamma^\dagger_{\mathbf{k}\sigma}c^\dagger_{\overline{\mathbf{q}}\overline{\sigma}}$ & 
$v^2_\mathbf{k}(1-f_{\mathbf{k}\sigma})
f_0(E_{\mathbf{k}\sigma}+eV)$ & -1 \\
\end{tabular}
\caption{Four terms in the tunnel Hamiltonian involving the state $\mathbf{k}\sigma$ in the superconductor.}\label{tab_tunneling}
\end{table}

For a simple qualitative model, we consider a BCS superconductor in high magnetic fields, including the effect of the Zeeman splitting \cite{fulde1973}, but neglecting orbital pair-breaking and spin-orbit scattering for simplicity. In the normal state, the electron energy relative to the chemical potential $\mu$ is given by $\epsilon_\mathbf{k}=\hbar^2\mathbf{k}^2/2m-\mu$. The quasiparticle energies in the superconducting state are $E_{\mathbf{k}\sigma}=E_\mathbf{k}+\sigma\mu_\mathrm{B}B$,
where $E_\mathbf{k}=\sqrt{\epsilon^2_\mathbf{k}+\Delta^2}$. Tunneling is described by a straightforward extension of \cite{tinkham1972b,zhao1995} to the case of finite Zeeman splitting: The tunnel Hamiltonian  for spin-conserving tunneling between states $\mathbf{k}\sigma$ and $\mathbf{q}\sigma$ in the superconductor and ferromagnet, respectively, is
\begin{equation}
H_T=\sum_{\mathbf{qk}\sigma} T_{\mathbf{qk}\sigma}c^\dagger_{\mathbf{k}\sigma}c_{\mathbf{q}\sigma}+H.c.
\end{equation}
Electron operators $c$ in the superconductor are written in terms of the quasiparticle and Cooper pair operators $\gamma$ and $S$ as
\begin{equation}
 c^\dagger_{\mathbf{k}\sigma}=u_\mathbf{k}\gamma^\dagger_{\mathbf{k}\sigma}+\sigma v_\mathbf{k}S^\dagger\gamma_{\overline{\mathbf{k}}\overline{\sigma}},
\end{equation}
where $\overline{\mathbf{k}}=-\mathbf{k}$, $\overline{\sigma}=-\sigma$, and the coherence factors are
\begin{equation}
 u^2_\mathbf{k}=\frac{1}{2}\left(1+\frac{\epsilon_\mathbf{k}}{E_\mathbf{k}}\right),\,\,\, 
 v^2_\mathbf{k}=\frac{1}{2}\left(1-\frac{\epsilon_\mathbf{k}}{E_\mathbf{k}}\right).
\end{equation}
The four terms appearing in the tunnel Hamiltonian 
for the state $\mathbf{k}\sigma$ in the superconductor are listed in table \ref{tab_tunneling}, where $f_{\mathbf{k}\sigma}$ is the quasiparticle distribution function in the superconductor, and $f_0$ is the Fermi function describing occupation in the ferromagnet. Summing up all contributions to the current for both spin directions yields
\begin{equation}
\begin{matrix}
I = & \displaystyle \frac{1}{e}\sum_{\sigma} \int \left[ G_\sigma u^2_\epsilon \left(f_{\epsilon\sigma}-f_0(E_{\epsilon\sigma}-eV)\right)\right. \\
& \displaystyle  - \left. G_{\overline{\sigma}} v^2_\epsilon\left(f_{\epsilon\sigma}-f_0(E_{\epsilon\sigma}+eV)\right)\right]d\epsilon,
\end{matrix}
\end{equation}
where we have replaced sums over $k$ by integration over $\epsilon$. We now group terms with equal energy $E$ inside ($\epsilon<0$) and outside ($\epsilon>0$) the Fermi surface, and introduce transverse and longitudinal distribution functions $f^{(T)}_{\epsilon\sigma}=(f_{\epsilon\sigma}-f_{\overline{\epsilon}\sigma})/2$ and $f^{(L)}_{\epsilon\sigma}=(f_{\epsilon\sigma}+f_{\overline{\epsilon}\sigma})/2$ in the superconductor. Setting $eV=0$ for the detector junction, we obtain
\begin{equation}
I = \frac{1}{e}\sum_{\sigma}\left[ \left( G_\sigma + G_{\overline{\sigma}}\right)Q^*_\sigma+
\left( G_\sigma -G_{\overline{\sigma}} \right) S_\sigma \right],\label{equ_Idet}
\end{equation}
where
\begin{eqnarray}
Q^*_\sigma & = & \int_{\epsilon>0} \left(u^2_\epsilon- v^2_\epsilon\right) f^{(T)}_{\epsilon\sigma} d\epsilon \\
S_\sigma & = & \int_{\epsilon>0} \left(f^{(L)}_{\epsilon\sigma}-f_0(E_{\epsilon\sigma})\right) d\epsilon
\end{eqnarray}
describe the charge imbalance and spin accumulation for each spin. The first term in (\ref{equ_Idet}) is the usual charge-imbalance signal observed in both the NISIN and FISIF sample. The second term is non-zero only for a ferromagnetic detector junction and gives rise to the observed peaks, as explained below.

Current injection in the bias regime of the Zeeman splitting creates both a charge imbalance $Q^*_\downarrow$ and spin accumulation $S_\downarrow$ in the lower Zeeman band, while the upper Zeeman band remains unoccupied. After diffusion over a distance $d>\lambda_{Q^*}$ along the superconductor we have $Q^*_\downarrow\approx 0$, but there is still a finite spin accumulation $S_{\downarrow}>0$ (spin-charge separation). The detector current is then
\begin{equation}
I_\mathrm{det} \approx\frac{1}{e}\left( G_\uparrow -G_\downarrow \right) S_{\downarrow}.
\end{equation}

We would now like to discuss the salient features of our experiment and some open questions within this model.
(i) We expect $S_\downarrow\propto |I_\mathrm{inj}|$, since for both bias directions there is an excess population of quasiparticles. Consequently, the sign of the detector current depends only on the sign of the spin polarization $P$. Since both $S_\downarrow$ and $I_\mathrm{det}$ are even functions of bias, the differential signals displayed in Figs.~\ref{fig_nonlocal} and \ref{fig_dist} are odd functions, explaining the asymmetric peaks. 
(ii) Once the upper Zeeman band is reached by the bias voltage, it yields an opposite contribution, canceling the signal. Therefore, the signal is restricted to the region of the Zeeman splitting, as seen in Fig.~\ref{fig_nonlocal}(c).
(iii) The peak height is proportional to the injector current. Therefore, the peak at negative bias is slightly larger than the peak at positive bias. The peak height ratio should be $(1+P)/(1-P)\approx 1.5$, which is consistent with the data shown in Figs.~\ref{fig_nonlocal}(a) and \ref{fig_dist}(a).
(iv) The normal-state spin diffusion length $\lambda_\mathrm{N}$ at low temperatures is determined by elastic spin flips due to spin-orbit scattering at nonmagnetic impurities (see \cite{zutic2004} and references therein for a comprehensive discussion). In superconductors without Zeeman splitting, the same mechanism is also expected to lead to spin relaxation (renormalized by coherence factors) \cite{yafet1983,*yamashita2002,*morten2004,*morten2005}. In the energy window of the Zeeman splitting, however, elastic spin flips can not relax the nonequilibrium spin accumulation $S_\downarrow$ due to the spin-dependent density of states. This explains why the observed spin relaxation length $\lambda_\mathrm{S}$ in the superconducting state at high fields is much larger than in the normal state. Weak relaxation paths might come from recombination of quasiparticles to Cooper pairs (possibly aided by spin-orbit scattering \cite{grimaldi1996}) or from inelastic spin flips to the upper Zeeman band. The latter would also explain the increase of $\lambda_\mathrm{S}$ with magnetic field, since larger energy transfer would be needed for increasing Zeeman splitting. A quantitative model for the relaxation mechanism is an open question to theory. A numerical simulation of the nonequilibrium distributions including a realistic model of inelastic scattering might also explain why the asymmetric nonlocal conductance signal extends to high bias in some of our samples. (v) Asymmetric peaks in the nonlocal conductance were also predicted for the competition of crossed Andreev reflection and elastic cotunneling in FISIF structures in the presence of Andreev bound states \cite{kalenkov2007,metalidis2010}. Bound states generated at the FIS interfaces \cite{huebler2012}, however, should extend no further than the coherence length into the superconductor, which is clearly inconsistent with the length scales observed in our experiment. Nevertheless, a common feature of bound states and Zeeman splitting is the existence of a spin-polarized density of states in the superconductor, and the conductance features look intriguingly similar. Another question to theory is therefore whether a coherent picture of both effects can be obtained. 

In conclusion, we have demonstrated spin injection and transport in superconductors in the regime of large Zeeman splitting of the density of states. We have found an asymmetric nonlocal conductance signal that can be modeled by spin accumulation in the lower Zeeman band of the quasiparticle dispersion. The relaxation length of the spin signal exceeds the normal-state spin-diffusion length by one order of magnitude, and we propose that it is determined by recombination of quasiparticles to Cooper pairs or inelastic spin flips. During preparation of this work we became aware of a similar study \cite{quay2012}.

We thank P. Machon, W. Belzig and M. Eschrig for useful discussions. This work was partially supported by the Deutsche Forschungsgemeinschaft under grant BE-4422/1-1. 

\bibliography{../../../lit.bib}

\end{document}